\documentstyle[aps,prd,amstex]{revtex}

\begin{document}
\title{Energy Localization Invariance of Tidal Work in General Relativity}

\author{Marc Favata\footnote{Present address: Department of Astronomy, Cornell University, Ithaca, NY 14853.}}
\address{Theoretical Astrophysics, California Institute of Technology,
Pasadena, CA 91125}

\date{printed \today}
\maketitle
\begin{abstract}
It is well known that,
when an external general relativistic (electric-type) tidal field ${\cal E}_{jk}(t)$
interacts with the evolving quadrupole moment ${\cal I}_{jk}(t)$ of an isolated body, 
the tidal field does work on the body (``tidal work'')---i.e., it transfers 
energy to the body---at a rate given by the same formula as in 
Newtonian theory: 
$dW/dt = -{1\over2}{\cal E}_{jk} d{\cal I}_{jk}/dt$.  Thorne has posed
the following question: In view of the fact that the gravitational interaction
energy $E_{\rm int}$
between the tidal field and the body is ambiguous by an amount $\sim
{\cal E}_{jk} {\cal I}_{jk}$, is the tidal work also ambiguous by this amount,
and therefore is the formula $dW/dt = -{1\over2}{\cal E}_{jk} d{\cal I}_{jk}/dt$
only valid unambiguously when integrated over timescales long compared
to that for ${\cal I}_{jk}$ to change substantially?  This paper completes
a demonstration that the answer is {\it no}; $dW/dt$ is {\it not} ambiguous in
this way.  More specifically, this paper shows that $dW/dt$ is unambiguously
given by $-{1\over2}{\cal E}_{jk} d{\cal I}_{jk}/dt$ independently of 
one's choice of how to localize gravitational energy in general relativity.
This is proved by explicitly computing $dW/dt$ using various gravitational 
stress-energy pseudotensors (Einstein, Landau-Lifshitz, M\o ller) as well as
Bergmann's conserved quantities which generalize many of the pseudotensors to include
an arbitrary function of position.  
A discussion is also given of the problem of
formulating conservation laws in general relativity and the role played by the various pseudotensors. 
\end{abstract}


\twocolumn
%\onecolumn
\narrowtext

\section{Introduction and Summary}
\label{sec:intro}

For many gravitating two body systems in the universe, the primary means of
energy transfer from one
body to the other is through tidal work. This work is accomplished
through the gravitational interaction between the tidal field of one body and
the mass multipole moments of the other body. A simple
example of this is the work that the moon does on the earth as it
raises the ocean's tides. Tidal work is also dramatically evident in 
the moon Io, which gets heated as it travels in an elliptical orbit through Jupiter's tidal 
gravitational-field. This heating is the cause of Io's dramatic volcanism. In these cases it is clear that the tidal work is a
physical observable and should in no way depend on one's means
of calculating it. 

The term ``tidal heating'' is often used in place of ``tidal work'', but is something of a misnomer. The net gravitational-energy that is transfered between
two bodies interacting tidally does not necessarily go into heat. It may go
into the energy needed to deform the body (i.e., raise a tide on it) or it may
go into the internal vibrational energy of the body. The net ``tidal work'' may
also be negative, in which case the phrase ``tidal cooling'' might be more
appropriate. Throughout this paper we will take the terms ``tidal heating'' and ``tidal work'' to be equivalent and to
mean the net work done by an external tidal field on an isolated body.

It seems evident that tidal work should be a ``physical observable''; i.e., the net energy-transfer from one body to another should be a real, physical quantity and
should not depend on the mathematics that one uses to calculate the work that
is done.

In calculating the tidal work for situations in general relativity, we
consider the interaction between an isolated body and a complicated ``external
universe'' (the precise definition of these terms will be discussed in Sec. III). In the most familiar cases, this external universe may simply refer
to a gravitating body such as a star, a planet, or a black hole, that orbits
around the isolated body. In such a system it has been shown by Thorne and
Hartle \cite{thornehartle} that the total mass-energy $M$ of the isolated body
is ambiguous by an amount $\Delta M \sim {\cal{I}}_{jk} {\cal{E}}_{jk}$, where
${\cal{I}}_{jk}$ is the mass quadrupole moment of the isolated body and
${\cal{E}}_{jk}$ is the tidal field of the external universe. 
This mass ambiguity has the same physical origin as the ambiguity in 
the localization of energy in a gravitational wave:  It arises from the
fact that there is no preferred way to localize gravitational energy.
This is true in Newtonian gravitational theory as well as in general relativity
\cite{purdue}.

This mass ambiguity shows up mathematically in
the fact that the nonlinearity of Einstein's equations could be expected
to produce a term $\sim {\cal E}_{jk} {\cal I}_{jk}/r$ in the time-time
component $g_{00}$ of the spacetime metric outside the body (where $r$ is a
radial coordinate); and one is free, mathematically, to move this term or 
some arbitrary part of it into the mass $M$ that appears in 
the standard equation $g_{00} = -1 + 2M/r + \ldots$.

We can also understand this mass-energy ambiguity 
physically in terms of the standard experiment by which
the total mass-energy $M$ of a gravitating body is measured: 
the application of 
the general relativistic version of Kepler's law 
to a test particle in orbit around the body.  If the body is
spherical and isolated and the orbit is circular, the body's mass-energy is 
related to the orbit's period $T$ (as measured by distant clocks) and
its radius $r$ (defined to be its circumference divided by $2\pi$) 
by $M= (r^3/G)(2\pi /T)^2$.  If the body is non-spherical, with
various multipole moments including the quadrupole moment ${\cal I}_{jk}$, 
then the moments perturb the
orbit; but if one makes the orbit as circular as those perturbations
permit and measures the orbit's average radius $\bar r$, then the $M$ 
that appears in the monopole $1/r$ part of the gravitational field 
is still given accurately, to first order 
in the moments, by the relativistic Kepler law 
$M= (\bar r^3/G)(2\pi /T)^2$.  The perturbations, being non-monopolar,
disappear when averaged over the orbit.
Similarly, if the body is precisely spherical but is
perturbed by a weak, external tidal field ${\cal E}_{jk}$, then accurate
to first order in those perturbations we can still compute $M$ by this
averaged-radius formula; again the perturbations average to zero over the
orbit (provided that the timescale on which ${\cal E}_{jk}$ changes is long compared to the orbital period). However, if {\it both} a quadrupole
moment ${\cal I}_{jk}$ and an external tidal field ${\cal E}_{jk}$ are   
present simultaneously, then the product ${\cal E}_{jk} {\cal I}_{jk}$ is 
monopolar in nature and has dimensions of mass; and correspondingly 
Kepler's law with an average radius will give $M+{\cal O}({\cal E}_{jk} {\cal
I}_{jk})$.  Thus, one cannot measure $M$ directly by Kepler's law.
We shall discuss this mass ambiguity further in 
Sec. \ref{sec:tidal}.

Zhang \cite{zhang1} has used the Landau-Lifshitz pseudotensor (one of
an infinite number of ways to localize gravitational-field energy) to derive
the expression 
\begin{equation}
{dW\over dt} = -{1\over{2}} {\cal E}_{jk} 
{d{\cal I}_{jk}\over dt}
\label{tidalwork}
\end{equation}
 for the
rate at which a time-evolving tidal field ${\cal E}_{jk}(t)$ does work on a body with 
time-evolving quadrupole moment ${\cal I}_{jk}(t)$.  In view of the body's
mass ambiguity $\Delta M \sim {\cal E}_{jk} {\cal I}_{jk}$, Zhang (and also
Thorne and Hartle \cite{thornehartle}) asserted that the work done
should be ambiguous by an amount $\sim \Delta M$, and thus Eq. (\ref{tidalwork})
should instead be written as $dW/dt = \langle -{1\over2} {\cal E}_{jk} d{\cal I}_{jk}/dt \rangle$ and  would be valid only
when averaged over timescales long enough for $W$ to build up by an amount
large compared to $\Delta M = {\cal E}_{jk}{\cal I}_{jk}$. This occurs, 
for example, in the long-term tidal heating of Io, during which
${\cal E}_{jk}$ and $d{\cal I}_{jk}/dt$ oscillate partially in phase with each other,
producing a cumulative work that goes into heat.

More recently Thorne \cite{thornetidal}, while analyzing
the effects of tidal forces on the stability of relativistic stars, 
claimed on physical grounds that Zhang \cite{zhang1} and Thorne and Hartle \cite{thornehartle} were wrong:  
The ambiguity $\Delta M$ actually resides 
solely in the energy of gravitational interaction $E_{\rm int}$ between the 
body and the external tidal field and not at all in the body's 
self energy $E_{\rm self}$ (i.e. the total mass-energy contained within the volume of the body), and correspondingly not at all in the work done 
by the tidal field on the body, $W = ($change in $E_{\rm self}$); so
the rate of work done is unambiguously and instantaneously given by
$dW/dt = -{1\over2}{\cal E}_{jk} d{\cal I}_{jk}/dt$ \footnote{Thorne \cite{thornetidal} needed this result as a key underpinning of his proof that tidal coupling stabilizes a star against gravitational collapse.}.

An operational variant of Thorne's argument is this:
Consider a body on which tidal work is being done by the interaction between its time-changing quadrupole moment ${\cal I}_{jk}(t)$ and some external time-changing tidal field ${\cal E}_{jk}(t)$. One can imagine, at any moment of time, turning off the tidal field 
while holding the body's size and shape unchanged (to first order in the 
tidal field).
With the tidal field gone, we can imagine measuring the body's total
mass-energy $M_{i}$, e.g., by the relativistic version of Kepler's laws.  That
measured mass-energy with tidal field (momentarily)
gone can be regarded as the 
body's self energy $E_{\rm self}$ (including rest mass).  This measured self energy
is unambiguous. Now, turn the tidal field back on and allow the system to evolve normally for some time $\Delta t$. Then, turn the tidal field off and make a second measurement of the body's total mass-energy $M_{f}$ in the same manner as before. The difference $M_{f} - M_{i}$ between these two measurements is the change in the body's self energy $\Delta E_{\rm self}$. This is the work $W$ done by the tidal field on the isolated body. We can then conclude that, since these measured changes in the body's self energy are unambiguous, $dW/dt = dE_{\rm self}/dt$ and $W$ are also unambiguous.

It is possible to test Thorne's claim in a manner based on the following
considerations:   
The self energy, defined in the above manner, will not change when the tidal
field ${\cal E}_{jk}(t)$ changes, but the shape and size
of the body are held fixed and thence ${\cal I}_{jk}$ is held fixed. This is just a restatement of the fact that a force can do no work if there is no displacement. However, if ${\cal I}_{jk}$ changes, with 
${\cal E}_{jk}$ held fixed, then  $E_{\rm self}$ {\it can} change.  This means that the unambiguous tidal work 
must be of the form $dW/dt = dE_{\rm self}/dt = ($some
constant$) \times {\cal E}_{jk} d{\cal I}_{jk}/dt$.  The interaction energy,
by contrast, should have the form of a product of the instantaneous tidal field
and quadrupole moment, so its time rate of change should be a perfect
time derivative $d E_{\rm int}/dt = d/dt[($some constant$)\times 
{\cal E}_{jk}{\cal I}_{jk}]$.  The body's total mass $M$ must be the sum
of its self energy and that portion of the interaction energy that resides
inside and near the body (i.e., within the orbit of the test particle that one uses in applying Kepler's third law to compute the mass), therefore $dM/dt = d E_{\rm self}/dt + dE_{\rm int}/dt = dW/dt
+ dE_{\rm int}/dt$.  If we express $dM/dt$ in the form 
\begin{equation}
{dM\over dt} = (\hbox{constant}) \times {\cal E}_{jk} 
{d{\cal I}_{jk}\over dt}  
+ (\hbox{constant})\times {d[{\cal E}_{jk}{\cal I}_{jk}]\over dt}\;,
\label{Decompose}
\end{equation}
then the first
term must be $dW/dt$ and the second $dE_{\rm int}/dt$.  If Thorne is
correct in his claim that $dW/dt$ is unambiguous and that the total ambiguity
of $dM/dt$ resides in $dE_{\rm int}/dt$, then {\it any} computation of
$dM/dt$ using any (general relativistically acceptable) localization of
gravitational energy must give $-1/2$ unambiguously for the coefficient 
of the first term, while different localizations should 
give different values for the coefficients of the second term.

Purdue \cite{purdue} has carried out detailed calculations of $dM/dt$ in
Newtonian theory using all possible localizations of the gravitational energy
and has found that, indeed, the first term in Eq.\ (\ref{Decompose})
always has the coefficient $-1/2$
while the second depends on the localization.  Purdue has also verified
that in general relativity, if one uses the energy localization embodied in the
Landau-Lifshitz pseudotensor, but performs gauge transformations (infinitesimal
coordinate transformations) on the spacetime metric,  the first coefficient
(that associated with $dW/dt$) remains always $-1/2$, while the second (that 
associated with $dE_{\rm
int}/dt$) changes with the changing gauge. 

In this paper we shall complete this test of Thorne's claim:  We shall
verify that, when one changes the general relativistic energy localization
by changing one's choice of pseudotensor, the first coefficient in Eq.\
(\ref{Decompose}) remains 
always $-1/2$ while the second changes and thus (partially) embodies the ambiguities present in localizing gravitational-field energy.

As a foundation for demonstrating this
we first discuss in Sec. \ref{sec:conserve} of
this
paper the problem of formulating covariant conservation laws in
general relativity. This is an underlying issue throughout this paper as the
lack of an acceptable energy-momentum tensor for the gravitational field
could possibly be a source of ambiguity in the calculation of the 
tidal work. We also discuss some of the various pseudotensors and conserved
quantities that are used to describe gravitational-energy localization.

In Sec. \ref{sec:tidal} we discuss the assumptions that go into our calculation
of the
tidal work. A key issue is that our calculations are performed in the
\emph{local asymptotic rest frame}, or LARF, of the body on which the work is being done. This means we are able to formulate only \emph{approximate}
conservation laws for our system. These laws are formulated in a \emph{buffer
zone} where the gravity of the isolated body is weak and the tidal field of the
external universe is uniform. The spacetime metric of this buffer zone is described in Sec. \ref{subsec:metric}.

We then calculate the tidal work using the Einstein pseudotensor
\cite{einstein} (Sec. \ref{subsec:tidaleinstein}) and review the calculation  given by Purdue \cite{purdue} using
the Landau-Lifshitz pseudotensor (Sec. \ref{subsec:tidalLL}). In Sec. \ref{subsec:tidalmoller} we perform the calculation using the
pseudotensor of M\o ller
\cite{moller}, which is significantly different from the two previously
mentioned pseudotensors, and in Sec. \ref{subsec:tidalbergmann} we examine the calculation using the conserved quantities
found by Bergmann \cite{bergmann}. Bergmann's conserved quantities generalize
many of the pseudotensors,
including those of Landau and Lifshitz, and Einstein. Each of these calculations gives the same, standard result $dW/dt = -{1\over2}{\cal E}_{jk} d{\cal I}_{jk}/dt$ for the tidal work, in agreement with Thorne's assertion.

Throughout this paper we adhere to the conventions of
Misner, Thorne, and Wheeler \cite{MTW}, hereafter
referred to as MTW. Space-time indices are represented by Greek letters and
spatial indices by Latin letters. We use units where $G=c=1$.
The constant on the right hand side of the Einstein field equations is
$+8\pi$ and the Minkowski flat-space metric $\eta_{\alpha \beta}$ has signature $(-,+,+,+)$.

\section{Conservation Laws and Pseudotensors}
\label{sec:conserve}
The formulation of covariant conservation laws has been a problematic
issue since general relativity's formulation in 1916. The issue has been
addressed by a large number of authors and some continue to work on this problem.

If one considers a system without gravitational fields, as in special
relativity, then the differential conservation laws for all matter and
energy fields present are given by the familiar formula
\begin{equation}
{T^{\mu \nu}}_{,\nu} = 0,
\label{eq:commaT}
\end{equation}
where $T^{\mu \nu}$ is the symmetric energy-momentum tensor of matter that
appears as a source term on the right-hand-side of the Einstein field 
equations
\begin{equation}
G^{\mu \nu}=8\pi T^{\mu \nu}.
\label{eq:efe}
\end{equation}
By matter we mean all fields with the exception of
the gravitational field.

In general relativity Eq. (\ref{eq:commaT}) is not an acceptable 
conservation law as it is not a tensor equation valid in all reference
frames. Instead we must use the covariant derivative in place of the
partial derivative and our equation becomes
\begin{equation}
{T^{\mu \nu}}_{;\nu}={T^{\mu \nu}}_{,\nu} + T^{\sigma \nu}
{\Gamma^{\mu}}_{\sigma \nu} + T^{\mu \sigma} {\Gamma^{\nu}}_{\sigma \nu}=0,
\label{eq:delT}
\end{equation}
where ${\Gamma^{\mu}}_{\sigma \nu}$ are the connection coefficients. From Eq. (\ref{eq:delT}) we can see that the mass-energy in matter fields is no
longer 
conserved as energy can now be transfered between the matter and the 
gravitational field. The quantity that is actually conserved  in the sense of Eq. (\ref{eq:commaT}) is some
\emph{effective} energy-momentum tensor $T_{\text{eff}}^{\mu \nu}$ of matter plus
gravitational fields which is given (in one variant) by Eq. (20.18) of MTW \cite{MTW} as  
\begin{equation}
T_{\text{eff}}^{\mu \nu}=T^{\mu \nu} + t^{\mu \nu},
\label{eq:Teff}
\end{equation}
where $t^{\mu \nu}$ is an energy-momentum pseudotensor for the
gravitational field. In other variants, some of which are  encountered below, $T_{\text{eff}}^{\mu \nu}=(-g)^{n/2} (T^{\mu \nu} + t^{\mu \nu})$ where $g=det\|g_{\alpha \beta}\|$ and $n$ is a positive integer\footnote{We will sometimes refer to scalars, vectors and tensors with factors of $(-g)^{n/2}$ in front as scalar, vector and tensor \emph{densities} of weight $n$.}. For each of these $T_{\text{eff}}^{\mu \nu}$, the equation ${T^{\mu \nu}}_{;\nu} = 0$ can be rewritten as 
\begin{equation}
T_{\text{eff} , \nu}^{\mu \nu}=0 ,
\label{eq:commaTeff}
\end{equation}
and $T_{\text{eff}}^{\mu \nu}$ can
be written as the divergence of some ``superpotential'' $H^{\mu [\nu \sigma]}$
that is antisymmetric in its last two indices \cite{LL}:
\begin{equation}
T_{\text{eff}}^{\mu \nu}={H^{\mu [\nu \sigma]}}_{,\sigma}.
\label{divsuper}
\end{equation}
Square brackets indicate antisymmetry of the tensor when the enclosed indices are swapped. Notice that Eq. (\ref{eq:commaTeff}) follows from Eq. (\ref{divsuper}) by differentiation and symmetry.

As mentioned above, $t^{\mu \nu}$ is not a true tensor, but
rather is a \emph{pseudotensor} that describes the localization of
gravitational energy-momentum. That $t^{\mu \nu}$ is not a tensor is a
fact intimately linked with Einstein's Equivalence Principle. Since we
are always free to choose our coordinates in spacetime to
correspond to a freely falling frame where the acceleration vanishes at a 
point, we
can equivalently choose a frame where the gravitational field vanishes
at that point. In
such a frame all the components of $t^{\mu \nu}$
% (for most pseudotensors)
 will likewise vanish at that point (provided one is using Minkowski
coordinates). However, in
any other reference frame, there is no reason why all the components of
$t^{\mu \nu}$ should vanish. Since any tensor that
vanishes in one reference frame must vanish in all reference frames, we can
conclude that $t^{\mu \nu}$ is not a tensor but a
pseudotensor and quantities calculated from it will depend on the choice of
one's
coordinate system. To make matters worse, $t^{\mu \nu}$ is defined only up to a
vanishing divergence, so there is an infinity of expressions for $t^{\mu \nu}$
corresponding to
an infinite number of ways in which one can localize the gravitational
energy-momentum density.

Despite their rather unpleasant nature in a theory so firmly rooted in
the principle of general covariance, pseudotensors have proved to be 
rather valuable calculational tools, especially in gravitational-wave
research (see for example \cite{thornekovacs}). The reason is that, despite their noncovariance, the $T_{\text{eff}}^{\mu \nu}$ can be used to compute covariant conserved quantities. For example, one can compute the total 4-momentum of a system that resides alone in asymptotically flat spacetime by the volume
integral
\begin{equation}
P^{\mu}=\int T_{\text{eff}}^{\mu 0}\, d^{3}x ,
\label{4momentum}
\end{equation}
where $d^{3}x=dx^{1}dx^{2}dx^{3}$ is a 3-volume element of constant time. Even though the integrand depends highly on one's choice of coordinates,
$P^{\mu}$ is a true vector that resides in the asymptotically flat region.

Using Gauss' law and the antisymmetry properties of the superpotential, it is also possible to express the 4-momentum as a surface integral:
\begin{equation}
P^{\mu}=\int {H^{\mu [0 \sigma]}}_{,\sigma} d^{3}x=\oint H^{\mu [0 j]} n_j d^{2}S ,
\label{Psurfint}
\end{equation}
where $n_j$ is the unit normal vector to the surface $S$. It is important to note that these integrals must be evaluated using an asymptotically Lorentz coordinate system\footnote{However, it should be noted that Nahmad-Achar and Schutz\cite{acharschutz} have devised a prescription for calculating pseudotensor-based conserved quantities for isolated systems in general relativity using coordinate systems with arbitrary asymptotic behavior.}.

\subsection{The Einstein Pseudotensor}
\label{sec:einstein}

The first pseudotensor was formulated by Einstein in 1916. The Einstein
pseudotensor is often referred to as the ``canonical'' pseudotensor because
it is derived using the general formula for the energy-momentum tensor of a
classical field with Lagrangian density $\cal{L}$ and field variable
$\eta_{A}$, which may be a tensor of any rank. In flat spacetime this general formula is given  by (see, e.g., Goldstein \cite{goldstein}),
\begin{equation}
T_{\mu \nu}={ {\partial \cal{L}}\over{\partial \eta_{A , \nu}} }
 \eta_{A , \mu} - \cal{L} \delta_{\mu \nu};
\label{eq:canonical}
\end{equation} 
and the Euler-Lagrange equations guarantee that ${T^{\mu \nu}}_{,\nu}=0$. In general relativity, the field variables are the components of the
metric tensor $g_{\mu \nu}$, and the Lagrangian density is given by
\begin{equation}
{\cal{L}} = {1 \over{16 \pi}} \sqrt{-g} g^{\alpha \beta} \left(
{\Gamma^{\gamma}}_{\alpha \beta}
 {\Gamma^{\sigma}}_{\gamma \sigma} - {\Gamma^{\sigma}}_{\alpha \gamma}
{\Gamma^{\gamma}}_{\beta \sigma} \right).
\label{lagrangian}
\end{equation}
Eq. (\ref{eq:canonical}) then becomes
\begin{equation}
\sqrt{-g} {}_{\text{E}}{t_{\mu}}^{\nu} =  \left( {{\partial \cal{L}}\over{\partial
g_{\alpha
\beta , \nu}}}  g_{\alpha \beta , \mu} - {\delta_{\mu}}^{\nu} \cal{L} \right),
\label{tE1}
\end{equation}
and again the Euler-Lagrange equations guarantee that $(\sqrt{-g} {}_{\text{E}}{t_{\mu}}^{\nu})_{,\nu}=0$ in vacuum and ${}_{\text{E}}{{{\mathfrak{T}}_{\mu}}^{\nu}}_{,\nu}=0$ where
\begin{equation}
{}_{\text{E}}{\mathfrak{T}_{\mu}}^{\nu}=\sqrt{-g} ({T_{\mu}}^{\nu} +
{}_{\text{E}}{t_{\mu}}^{\nu})
\label{einsteincomplex}
\end{equation}
when matter is present. The tensor density ${}_{\text{E}}{\mathfrak{T}_{\mu}}^{\nu}$ is often referred to as a ``total energy-momentum complex''; it is the Einstein variant of the $T_{\text{eff}}^{\mu \nu}$ discussed above.

From Eq. (\ref{tE1}) we arrive at an explicit expression for the Einstein
pseudotensor \cite{dirac}:
\begin{multline}
\sqrt{-g} {}_{\text{E}}{t_{\mu}}^{\nu}= \\
{1 \over{16 \pi} } \left( \left( {\Gamma^{\nu}}_{\alpha \beta} -
\delta_{\beta}^{\nu} {\Gamma^{\sigma}}_{\alpha \sigma} \right) \left(g^{\alpha
\beta} \sqrt{-g} \right)_{,\mu} - \delta^{\nu}_{\mu} \cal{L} \right).
\label{tE2}
\end{multline}
Note that raising or lowering an index for this pseudotensor does not
produce a symmetric quantity, so we are unable to form a
conserved angular-momentum complex from it.

It was shown by von Freud \cite{vonfreud} that the Einstein complex can be
written as the divergence of an antisymmetric ``superpotential''
${}_{\text{F}}U_{\alpha}^{[\beta \gamma]}$:
\begin{equation}
{}_{\text{E}}{\mathfrak{T}_{\mu}}^{\nu} = {}_{\text{F}} {U^{[\nu \sigma]}_{\mu}}_{,\sigma}
\label{divvonfreudsuper}
\end{equation}
where 
\begin{multline}
{}_{\text{F}}U_{\alpha}^{[\beta \gamma]}= \\
- {1 \over{16 \pi} } {g_{\alpha \sigma} \over{\sqrt{-g}}} \left\{ -g \left(
g^{\beta \sigma} g^{\gamma \lambda} - g^{\gamma \sigma} g^{\beta \lambda}
\right) \right\}_{,\lambda} .
\label{vonfreudsuper}
\end{multline}

We can now form expressions for the covariant components of the 4-momentum of an isolated system by means of Eqs. (\ref{4momentum}) and (\ref{Psurfint}):
\begin{equation}
P_{\mu}= \int {}_{\text{E}}{\mathfrak{T}_{\mu}}^{0} \: d^{3}x=\oint {}_{\text{F}}U_{\mu}^{[0 j]}
n_j d^{2}S ,
\label{einsteinint}
\end{equation}
where the first integral is over the system's entire volume, and the second is over a closed surface near spatial infinity (in the asymptotically flat region of spacetime).
Because of the peculiarities of the pseudotensor, the
above integral can only be interpreted as the covariant components of an
energy-momentum 4-vector if one is using coordinates
$x^{\mu}=(t,x,y,z)$ in which the metric $g_{\alpha \beta}$ asymptotically
approaches the Minkowski flat metric $\eta_{\alpha \beta}$. 

To illustrate the coordinate dependent nature of the pseudotensors, we
provide two well known examples (mentioned by M\o ller \cite{moller}  
and Anderson \cite{anderson}). If one were to calculate the integral $\int
 {}_{\text{E}}{t_{0}}^{0} \: d^3 x$ for the Lorentz metric $g_{\alpha \beta}=\eta_{\alpha
\beta}$, its value would be zero, the expected energy for a region with no
gravitational field. However, if we merely change to spherical coordinates,
the value of this integral is infinite even though spacetime is flat.
Similarly, evaluation
of the integral for the Schwarzschild metric only yields the mass $M$ if
one uses coordinates such that $g_{\alpha \beta}$ maps to $\eta_{\alpha
\beta}$ asymptotically as $r \to \infty$.

Despite the restrictions on the use of this pseudotensor, it has still led
to the reliable prediction by Einstein that gravitational waves exist and carry
a definite energy.

\subsection{The Landau-Lifshitz Pseudotensor}
\label{sec:LL}

Landau and Lifshitz (LL) \cite{LL} were able to formulate a symmetric
pseudotensor, thus allowing the construction of a conserved total
angular-momentum complex. Their conserved total energy-momentum complex (their variant of $T_{\text{eff}}^{\mu \nu}$) is given by
\begin{equation}
{\frak{T}}_{\text{LL}}^{\mu \nu}={h^{\mu [\nu \sigma]}}_{,\sigma} = (-g) (T^{\mu
\nu} + t_{\text{LL}}^{\mu \nu})
\label{totalLL}
\end{equation}
(where $h^{\mu [\nu \sigma]}$ is defined below) and satisfies the usual property
\begin{equation}
{{\frak{T}}_{\text{LL}}^{\mu \nu}}_{,\nu} = 0.
\label{TLLconserved}
\end{equation}
The explicit form of $t_{\text{LL}}^{\mu \nu}$ is long and complicated. It is given by 
\begin{align}
(-g) t_{\text{LL}}^{\mu \nu} &= \text{[Eq. (20.23) of MTW]}\\
 &= \text{[Eq. (96.9) of LL]}.
\label{tLL}
\end{align}

The LL superpotential $h^{\mu [\nu \sigma]}$ is related to that given by von
Freud by \cite{anderson}:
\begin{equation}
h^{\mu [\nu \sigma]}=\sqrt{-g} g^{\mu \rho} {}_{\text{F}}U_{\rho}^{[\nu \sigma]}.
\label{LLvonfreud}
\end{equation}

The LL pseudotensor is related to the Einstein pseudotensor by the following formula \cite{anderson}:
\begin{equation}
(-g)t_{\text{LL}}^{\mu \nu}=(-g)g^{\mu \rho} {}_{\text{E}} {t_{\rho}}^{\nu}+(\sqrt{-g}g^{\mu \rho})_{,\sigma} {}_{\text{F}} {U_{\rho}}^{[\nu \sigma]}.
\label{relateLLtoE}
\end{equation}

As in the case with the Einstein pseudotensor, integrals of the Landau-Lifshitz pseudotensor also produce strange results in curvilinear coordinate systems. Asymptotically Lorentz coordinates must again be used if one wants sensible results.

\subsection{The M\o ller Pseudotensor}
\label{sec:moller}

The M\o ller pseudotensor \cite{moller} is significantly different from 
the two complexes mentioned above. In deriving his pseudotensor M\o ller
sought to eliminate the problem that the integral given in Eq.
(\ref{einsteinint}) yields strange results if one converts to curvilinear
coordinate systems. 

To define his pseudotensor, M\o ller makes use of the fact that one can
always add a quantity ${S_{\mu}}^{\nu}$ to the Einstein complex and still
retain energy-momentum conservation, provided that
${{S_{\mu}}^{\nu}}_{,\nu}=0$. The new total pseudotensor complex (matter plus
gravitational fields) will thus have the form
\begin{equation}
{}_{\text{M}}{\mathfrak{T}_{\mu}}^{\nu} = {}_{\text{E}}{\mathfrak{T}_{\mu}}^{\nu} +
{S_{\mu}}^{\nu}.
\label{moller58}
\end{equation}
M\o ller additionally restricts the form of ${}_{\text{M}}{\mathfrak{T}_{\mu}}^{\nu}$ by
requiring the following conditions (see also Komar \cite{komar}):

1. It must be identically conserved: ${}_{\text{M}}{{\mathfrak{T}_{\mu}}^{\nu}}_{,\nu}=0$.

2. The integral over some 3-volume of constant time of ${}_{\text{M}}\mathfrak{T}_{\mu}
\, ^{\nu}$ must produce the same results (in an asymptotically Lorentz coordinate system) as Eq. (\ref{einsteinint}):
\begin{equation}
\int {}_{\text{M}}{\mathfrak{T}_{\mu}}^{0} \, d^{3}x = \int {}_{\text{E}}{\mathfrak{T}_{\mu}}^{0}
\, d^{3}x.
\label{moller59}
\end{equation}

3. ${}_{\text{M}}{{\mathfrak{T}}_{0}}^{0}$ and ${}_{\text{M}}{{\mathfrak{T}}_{0}}^{\nu}$ behave like scalar and vector densities under
arbitrary changes of the \emph{spatial} coordinates, $x^{j}_{\rm{new}}=F^{j}(x^{1}_{\rm{old}},x^{2}_{\rm{old}},x^{3}_{\rm{old}})$, $x^{0}_{\rm{new}}=x^{0}_{\rm{old}}$. This allows one to change the
coordinate system from say, Minkowski to spherical, but not to change the
way one slices spacetime into space plus time.

4. Under \emph{linear} transformations, ${}_{\text{M}}{\mathfrak{T}_{\mu}}^{\nu}$ behaves like a mixed second-rank tensor.

With the further restriction that it not contain higher than second order derivatives of the metric, M\o ller explicitly exhibits a unique energy-momentum complex with these properties, and he shows that it can be written as the divergence of the following antisymmetric superpotential ${\chi_{\mu}}^{[\nu
\sigma]}$:
\begin{multline}
{}_{\text{M}}{\mathfrak{T}_{\mu}}^{\nu} = {{\partial \chi_{\mu}^{[\nu
\sigma]}}\over{\partial x^{\sigma}}} = \\
 - {1\over{8\pi}} {\partial \over{\partial x^{\sigma}}} \big[\sqrt{-g} (g_{\mu
\alpha , \beta} - g_{\mu \beta , \alpha}) g^{\nu \beta} g^{\sigma \alpha}\big].
\label{mollerpseudotensor}
\end{multline}

The M\o ller superpotential $\chi_{\mu}^{[\nu \sigma]}$ is related to the von
Freud superpotential by \cite{anderson}, \cite{komar}:
\begin{equation}
\chi_{\mu}^{[\nu \sigma]}=2 {}_{\text{F}}U_{\mu}^{[\nu \sigma]} - \delta_{\mu}^{\nu}
{}_{\text{F}}U_{\rho}^{[\rho \sigma]} + \delta_{\mu}^{\sigma} {}_{\text{F}}U_{\rho}^{[\rho \nu]}
\label{mollervonfreud}
\end{equation} 

Like the Einstein pseudotensor, the M\o ller complex is not symmetric and thus
cannot be used to form conservation laws for angular momentum. Moreover, unlike
the complexes of Einstein and Landau-Lifshitz, the M\o ller complex is not
entirely quadratic in the first derivatives of the metric but has a term that
is linear in the second derivatives of the metric. As pointed out by M\o ller himself \cite{moller}, this means that ${}_{\text{M}}{\mathfrak{T}_{\mu}}^{\nu}$ will generally \emph{not} vanish in a local Lorentz frame with no matter present.
This will become an issue in Sec. \ref{sec:tidal},  when we use the M\o ller complex to calculate the tidal work.

\subsection{The Bergmann Conserved Quantities}
\label{sec:berg}

Recognizing that conservation laws are related to the invariance properties of physical laws, and combining this with the fact that the equations of general relativity are invariant under arbitrary coordinate transformations, Bergmann \cite{bergmann} proposed that to each infinitesimal coordinate transformation there would correspond a conserved quantity.
Making use of various identities, Bergmann \cite{bergmann} constructs a relationship between an arbitrary vector field $\xi^{\sigma}$ (which may be thought of as inducing infinitesimal coordinate transformations on the metric) and the generators $C^{\rho}$ of these transformations:
\begin{equation}
\sqrt{-g}G^{\mu \nu} \delta g_{\mu \nu}+{C^{\rho}}_{,\rho} \equiv 0
\label{berg2}
\end{equation}
where 
\begin{equation}
\delta g_{\mu \nu}=-(\xi_{\mu ; \nu}+\xi_{\nu ; \mu}).
\label{metrictransform}
\end{equation}
Bergmann chooses $C^{\rho}=2\sqrt{-g}G^{\rho \sigma} \xi_{\sigma}$ as a solution to Eq. (\ref{berg2}). However, one may always add an arbitrary curl field ${V^{[\rho \sigma]}}_{,\sigma}$ to $C^{\rho}$ and still satisfy Eq. (\ref{berg2}). Bergmann chooses this curl such that the resulting expression contains no higher than first derivatives of the metric. His final expression satisfying Eq. (\ref{berg2}) is
\begin{equation}
\bar{C}^{\mu}= 2 \xi^{\sigma} \sqrt{-g} {G_{\sigma}}^{\mu} +
{\left(\xi^{\sigma} {}_{\text{F}}U_{\sigma}^{[\mu \nu]} \right)}_{,\nu}.
\label{berg1}
\end{equation}
Eq. (\ref{berg1}) represents a weakly conserved quantity, meaning that it satisfies ${\bar{C}^{\mu}}_{\; \, \;,\mu}=0$ whenever the vacuum field equations are satisfied ($G^{\mu \nu}=0$). The corresponding strong conservation law is ${D^{\mu}}_{,\mu}=0$, where
\begin{equation}
D^{\mu}\equiv {\left( \xi^{\sigma} {}_{\text{F}}U_{\sigma}^{[\mu \nu]} \right)}_{,\nu} \equiv \bar{C}^{\mu}-2\sqrt{-g}{G_{\sigma}}^{\mu} \xi^{\sigma},
\label{strongconserve}
\end{equation} 
as can be easily shown if we
make use of the antisymmetry of ${}_{\text{F}}U_{\sigma}^{[\mu \nu]}$ and the
commutativity of partial derivatives.

From this strongly conserved quantity Bergmann constructs the 4-momentum in the same manner as we did from Eq. (\ref{4momentum}):
\begin{equation}
P^{\mu}=\int D^{\mu} d^{3}x.
\label{bergint}
\end{equation}
Using Gauss' law and the antisymmetry of ${}_{\text{F}}U_{\sigma}^{[\mu \nu]}$, we can write the $\mu=0$ component as a surface integral,
\begin{equation}
P^{0}= \oint \xi^{\sigma} {}_{\text{F}}U_{\sigma}^{[0 j]} n_j d^{2}S.
\label{bergsurfint1}
\end{equation}

According to Bergmann, from the weakly conserved quantities $\bar{C}^{\mu}$, expressions equivalent to
several of the pseudotensors can be derived by making specific choices for
$\xi^{\sigma}$. For example choosing $\xi^{\sigma}=k^{\sigma}$ yields the canonical-Einstein expression contracted with $k^{\sigma}$, while setting $\xi^{\sigma}=\sqrt{-g} g^{\sigma \alpha} k_{\alpha}$
yields the Landau-Lifshitz expression contracted with $k^{\sigma}$. 
We have been unable to find a similar choice that yields the M\o ller pseudotensor. We believe that this is due to Bergmann's choice of the curl field ${V^{[\rho \sigma]}}_{,\sigma}$ containing no second derivatives of the metric. Recall that unlike most other pseudotensors, the M\o ller complex contains second derivatives of the metric.

\subsection{Other Formulations of the Conservation Laws}
\label{sec:otherconserve}

Aside from the method of using pseudotensors to formulate conservation laws in
general relativity, there exist several other approaches as well. One of these
is the method of \emph{quasilocal} energy, a covariant definition of energy that arises from a Hamiltonian formulation of general relativity.
For a discussion of the equivalence of the quasilocal and pseudotensor approaches
to gravitational energy-momentum, see Chang, Nester, and Chen
\cite{changnester}. In parallel with this present research, Booth and Creighton \cite{jolien} have calculated the tidal work using the Brown and York \cite{brownyork} quasilocal energy approach, and have arrived at the same result as we deduce below using various pseudotensors: $dW/dt = -{1\over2}{\cal E}_{jk} d{\cal I}_{jk}/dt$.

There have also been efforts to find an energy-momentum
\emph{tensor} for the gravitational field. Shortly after relativity was
formulated, Lorentz and Levi-Civita proposed that the Einstein tensor $G^{\mu
\nu}$ be used as a gravitational energy-momentum tensor. This however, did not
prove fruitful \footnote{For an excellent discussion of the exchange that
occurred between Einstein, Levi-Civita, Lorentz, and others concerning
conservation laws and the prediction of gravitational waves, see the article by
Cattani and De Maria \cite{cattani}}. Most recently, Babak and Grishchuk
\cite{grishchuk} have shown that when one formulates general relativity as a non-linear field theory in flat spacetime, then there exists an energy-momentum tensor for the gravitational field that has all the nice properties one might wish. This energy-momentum tensor is the Landau-Lifshitz
pseudotensor with partial derivatives replaced by covariant derivatives with
respect to the flat-background metric.

\section{Calculation of the Tidal Work}
\label{sec:tidal}

In calculating the tidal work, we consider a system consisting of an
\emph{isolated} body that interacts with a complicated external universe in the
\emph{slow-motion} approximation. The body is isolated in the sense that the radius of
curvature $\cal{R}$ of the external universe and the lengthscale $\cal{L}$ on
which this curvature changes must both be large when compared with the size $R$
of the isolated body: $R/{\cal{R}} \ll 1$ and $R/{\cal{L}} \ll 1$. This means
that the external universe is not subjecting the isolated body to very strong
gravitational fields (as would happen e.g., in a neutron star and black hole close to
merger) and that the tidal field of the external universe is nearly uniform in the
region near the isolated body. By slow-motion, we mean that the timescale
$\tau$ for changes in the mass and current moments of the body and the tidal field of the external universe
are small compared to the size of the body: ${R/{\tau}}  \ll 1$. If this were not the case, we would have to worry about changes in
the mass-energy $M$ due to gravitational radiation and other higher-order
effects. For detailed discussions of the constraints $R/{\cal{R}} \ll 1$, $R/{\cal{L}} \ll 1$, and ${R/{\tau}}  \ll 1$, and of various approximations based on them (which we shall use below), see Thorne and Hartle \cite{thornehartle} and the recent paper by Purdue \cite{purdue}, whose analysis we are continuing.

 Some examples of isolated, slow-motion bodies discussed by Purdue \cite{purdue}
include: (i) a compact object such as a neutron star or black hole in a binary
inspiraling system that is not too close to merger; and (ii) Jupiter's moon Io, which
gets tidally heated as it travels through Jupiter's tidal field in an
elliptical orbit. 

Our calculation of the tidal work involves computing $dM/dt$, the rate of change of the mass of an isolated body, and then expressing $dM/dt$ in the form of Eq. (\ref{Decompose}) and reading off the two coefficients. We use the multipole moment formalism discussed in Thorne \cite{thornemultipole} and Thorne and Hartle \cite{thornehartle} and treat gravity
as a non-linear field theory in flat spacetime. The computation of $dM/dt$ is
carried out as a 2-dimensional surface integral of a pseudotensor in the
``buffer zone''or local asymptotic rest frame (LARF) of the isolated body (see
Eq. (2.3a) of Thorne and Hartle \cite{thornehartle}):
\begin{equation}
{{dM}\over{dt}}=- \oint t^{0j} \, d^{2}S_j , 
\label{surfaceint}
\end{equation}
where $d^{2}S_j=n_j r^{2} d\Omega$ is the surface element of a 2-sphere $\partial \cal{V}$ in the buffer zone that encloses a volume $\cal{V}$ and has unit normal $n_j$ and solid angle $d\Omega$. This buffer zone is a region that surrounds the isolated body but is far enough
away that gravity in it can be considered weak. At the same time, it is close
enough to the body that the tidal field of the external universe appears
homogeneous. The buffer zone can be described more precisely \cite{purdue} as
the region where ${r/{\cal{L}}} \ll {1}$, $r/{{\cal{R}}} \ll {1}$, and $M/r \ll
1$, $r$ being the radial distance from the isolated body.  The rate of change of mass-energy through the surface $\partial \cal{V}$ is $dM/dt$,  and $M$ is the total mass-energy inside $\cal{V}$.
Note that our analysis is thus valid even for a strongly gravitating body such as a black hole, provided there exists a buffer region around it where gravity is weak, the external curvature is nearly uniform,  and the spacetime curvature is not changing too rapidly.

For the purposes of our discussion and to the order that our calculations are
valid, there are only three relevant parameters that characterize the spacetime:\\
 1. The total mass-energy $M$ of the isolated body.\\
2.  The quadrupole moment ${\cal{I}}_{jk}$ of the isolated body, which, in the
limit of weak gravity, is given by
\begin{equation}
{\cal{I}} _{jk} = \int \rho x_{j} x_{k} - {1 \over{3}} \delta_{j k} r^{2} \:
d^{3}x .
\label{quadrupole}
\end{equation}
3. The tidal field ${\cal{E}} _{jk} = R_{j 0 k 0}$ of the external universe,
where $R_{\alpha \beta \gamma \delta}$ is the Riemann tensor of the external
universe.

Note that both ${\cal{I}}_{jk}$ and ${\cal{E}}_{jk}$ are symmetric and trace
free tensors that reside in the buffer zone, and that we are using coordinates that are as Lorentz as possible (with respect to the physical metric) throughout the buffer zone; i.e., $g_{\alpha \beta}=\eta_{\alpha \beta}+O(M/r)+O({\cal{I}}/r^3)+O({\cal{E}}r^2)$. These ${\cal{I}}_{jk}$ and ${\cal{E}}_{jk}$ are spatially constant in the buffer region but they may depend on time.

The body has additional multipole moments: the current quadrupole moment ${\cal{S}}_{jk}$, the mass octupole moment ${\cal{I}}_{jkl}$, etc; and the external universe has additional tidal fields of ``magnetic-type'' (${\cal{B}}_{jk}$, \ldots) and ``electric-type'' (${\cal{E}}_{jkl}$, \ldots), see e.g., Thorne and Hartle \cite{thornehartle}. These moments and tidal fields can couple to each other to produce tidal work: $dW/dt \sim { {\cal{B}}_{jk}} {d {\cal{S}}_{jk}}/{dt} \;\&\; { {\cal{E}}_{jkl}} {d {\cal{I}}_{jkl}}/{dt} \;\&\; \cdots$. In some situations these contributions to $dW/dt$ might be larger than the one, $dW/dt \sim { {\cal{E}}_{jk}} {d {\cal{I}}_{jk}}/{dt}$, that we are studying, but typically the mass quadrupole will dominate. In this paper we restrict ourselves to the mass quadrupole term, whether or not it dominates the tidal work, because we are seeking to discuss an issue of principle first raised by Thorne and Hartle \cite{thornehartle}: the non-ambiguity of the ${ {\cal{E}}_{jk}} {d {\cal{I}}_{jk}}/{dt}$ tidal work. Presumably our results can be generalized to higher order moments, but we shall not attempt to do so. Correspondingly, in our analysis we shall consider only $M$, ${\cal{I}}_{jk}$, and ${\cal{E}} _{jk}$.

Keep in mind that when we identify a mass $M$ as in Eq. (\ref{surfaceint}), we
are only doing so in an approximate sense. This is because the mass, momentum,
and angular momentum only have precisely defined values in an asymptotically
flat spacetime. As our spacetime is only \emph{locally} asymptotically flat,
the conservation laws only give approximate values of the mass, momentum, and
angular momentum in the buffer zone where spacetime is \emph{approximately}
asymptotically flat.

In particular, as we discussed in Sec. \ref{sec:intro}, there is an ambiguity in the mass $M$ of the
isolated body $\Delta M \sim {\cal{I}}_{jk} {\cal{E}}_{jk}$. If ${\cal{I}}_{jk}$ is oscillating in time,  then this ambiguity is
of the same order as the amount of energy that is transfered between the
isolated body and the external universe by tidal work during one period of
oscillation. We can understand
this ambiguity more clearly by examining the time-time part of the metric in
the buffer zone \cite{purdue}:
\begin{equation}
g_{0 0}=-1 + {2M\over{r}} + 3 {\cal{I}}_{ab} {{x^{a} x^{b}} \over{r^{5}}} +
\cdots - {\cal{E}}_{ab} x^{a} x^{b} + \cdots  ,
\label{g00}
\end{equation}
 where $r$ is the distance from the center of the isolated body as measured in
its local asymptotic rest frame. We have omitted terms involving higher order mass
and tidal multipole moments (e.g. octupole moments ${\cal{I}}_{jkl}$ and
${\cal{E}}_{jkl}$) and also terms that are products of $M$, ${\cal{I}}_{jk}$,
and ${\cal{E}}_{jk}$ which result from the nonlinearities of the
Einstein field equations. One of these nonlinear terms has the form
\begin{equation}
{\delta g_{0 0}} \sim { {{\cal{I}}_{jk} {\cal{E}}_{jk}}\over{r}}
\label{ambig}
\end{equation}
which has the same form as $2M/r$, i.e. $({\rm{monopole}})/r$, and which has a coefficient
that is gauge dependent. The similarity in structure between $M/r$ and $({\cal{I}}_{jk} {\cal{E}}_{jk})/r$ implies that it is
possible to move portions of the gauge-dependent term given in Eq. (\ref{ambig})
into or out of the $2M/r$ term. One can interpret this as meaning that the mass
$M$ that one reads off the metric is ambiguous by an amount on the order of 
\begin{equation}
\Delta M \sim {\cal{I}}_{jk} {\cal{E}}_{jk}.
\label{deltam}
\end{equation}

Purdue \cite{purdue} shows that this ambiguity is also present in Newtonian theory
in the form of an ambiguous gravitational interaction-energy inside and near the body. More specifically: The total mass-energy $E_{\cal{V}}$ enclosed in the volume $\cal{V}$ can be
expressed in Newtonian theory as
\begin{equation}
E_{\cal{V}}=E_{\text{self}} + E_{\text{e}} + E_{\text{int}} ,
\label{EVol}
\end{equation}
where $E_{\text{self}}$ is the isolated body's self-energy (which depends on the body's rest-mass and internal energy density distributions), $E_{\text{e}}$ is the
external field energy inside the volume ${\cal{V}}$(which depends only on the external tidal field
${\cal{E}}_{jk}$), and $E_{\text{int}}$ is the interaction energy inside ${\cal{V}}$ and is given by
\begin{equation}
E_{\text{int}}=\left( {{2+\alpha} \over 10} \right) {\cal{I}}_{jk}
{\cal{E}}_{jk}.
\label{eq:eint}
\end{equation}
Here the coefficient $\alpha$ depends on one's choice of Newtonian
energy-localization. For example (see Purdue \cite{purdue}), the choice $\alpha=0$ localizes all of the
gravitational energy in the field, so that the total gravitational energy is
given by a volume integral of $(\nabla \Phi)^2 /(8 \pi)$ (which should be
familiar from electrostatics). Alternatively, the choice $\alpha=1/2$ localizes
the gravitational energy entirely in the matter, so the total-gravitational
energy is given by a volume integral of ${1\over{2}} \rho \Phi$. Here $\rho$
is the mass density, so clearly it vanishes outside the material of a
gravitating body. 

From Eq. (\ref{EVol}) we can see that a Newtonian
calculation of $dE_{\cal{V}}/dt$ (which is analogous to our
general-relativistic calculation of $dM/dt$) will include a term that is the total
time derivative of the interaction term $E_{\text{int}}$. Our
general-relativistic calculation will also have an interaction term of this
same form, where the coefficient in front depends on the energy-localization
scheme. In the Newtonian case, Purdue shows that, despite the ambiguity of $dE_{\text{int}}/dt$, the rate of change of the body's self energy $dE_{\text{self}}/dt$ is given unambiguously by the tidal work formula $dE_{\text{self}}/dt = -{1\over2}{\cal E}_{jk} d{\cal I}_{jk}/dt$. Our general relativistic analysis will produce this same conclusion.

\subsection{Metric in the Buffer Zone}
\label{subsec:metric}

In our calculation we will consider general relativistic gravity not as a
geometric phenomenon involving the curvature of spacetime, but rather as a
non-linear field theory in flat spacetime. We treat the field variables
(the metric components $g_{\alpha \beta}$) as a perturbative expansion in some
dimensionless parameter $\varepsilon$ which is actually the gravitation
constant, $G=1$  in our system of units. Thus, terms of $O(\varepsilon)$ are
linear perturbations around flat spacetime; terms of $O(\varepsilon^{2})$ are
quadratic, etc. All raising and lowering of indices is done with the flat Minkowski metric $\eta_{\alpha \beta}$. 

Our three parameters $M$, ${\cal{I}}_{jk}$, and ${\cal{E}}_{jk}$ can all be
considered linear in $\varepsilon$. In our calculation of the tidal work, it
is clear from the form of Eq. (\ref{Decompose}) that we may also need to consider
terms in the metric that are quadratic in $\varepsilon$, as such terms may go
as ${\cal{I}}_{jk} {\cal{E}}_{jk}$. We thus expand the metric up to quadratic
order:
\begin{equation}
g_{\alpha \beta}=\eta_{\alpha \beta}+\varepsilon h_{\alpha \beta}+\varepsilon^2
k_{\alpha \beta},
\label{eq:metric}
\end{equation}
where $h_{\alpha \beta}$ contains terms that are linear in $M$,
${\cal{I}}_{jk}$, and ${\cal{E}}_{jk}$ and their time derivatives, while
$k_{\alpha \beta}$ contains terms that are products of any two of those three
quantities and their first time derivatives (for example $M \cal{I}$, $\dot{\cal{I}}
\cal{I}$, $\cal{I} \cal{E}$, $\dot{\cal{E}} \dot{\cal{E}}$, etc). Any terms
that are cubic or higher in the perturbation expansion $\varepsilon$ cannot
contribute to the tidal work (which is itself of order $\varepsilon^{2}$) and can be discarded at any point in the
calculation. 

Since the mass quadrupole moment ${\cal{I}}_{jk}$ and the tidal
field ${\cal{E}}_{jk}$ are spatially uniform in the buffer zone, the spatial
gradients of these functions vanish. Furthermore, since $\partial_t \sim 1/\tau$, the slow-motion approximation allows us to ignore all second and higher order time derivatives of ${\cal{I}}_{jk}$ and ${\cal{E}}_{jk}$.

We impose separate gauge conditions on the linear $h_{\alpha \beta}$ and
quadratic $k_{\alpha \beta}$ parts of the metric. The linear part of the metric
has been calculated by Zhang \cite{zhang1} in the de Donder gauge and is used
by Purdue \cite{purdue} in her analysis. The de Donder gauge in linear
order\footnote{The general de Donder condition is given by ${{\frak{g}}^{\mu
\nu}}_{,\nu}=0$ where ${\frak{g}}^{\mu \nu}=\sqrt{-g} g^{\mu \nu}$} is given by
the condition that ${{\bar{h}}^{\mu \nu}}_{\; \: \; \; ,\nu}=0$, where $\bar{h}^{\mu \nu}$ is
the trace reversed metric perturbation:
\begin{equation}
\bar{h}^{\mu \nu}\equiv h^{\mu \nu} - {1 \over{2}} \eta^{\mu \nu} h.
\label{hbar}
\end{equation}
In terms of the metric perturbation $h_{\alpha \beta}$ we can write this as the
condition:
\begin{equation}
{h^{\mu \nu}}_{,\nu}={1 \over{2}}h^{,\mu} \;,
\label{dedonder}
\end{equation}
where $h$ is the trace of $h_{\alpha \beta}$:
\begin{equation}
h={h_{\alpha}}^{\alpha}=\eta^{\alpha \beta} h_{\alpha \beta} .
\label{traceh}
\end{equation}

To the order we should need in the slow-motion approximation (i.e. neglecting second and higher order time derivatives) and ignoring higher order
multipole moments (i.e. ${\cal{I}}_{jkl},\, {\cal{S}}_{jk},\, {\cal{E}}_{jkl},\, {\cal{B}}_{jk}$, etc.) the linear part of the metric is given in Cartesian
coordinates and de Donder gauge, accurate to $O(\varepsilon)$ in the buffer zone, by \cite{zhang1}:
\begin{align}
h_{00} &\equiv -2\Phi = 2 {M \over r} + 3 {{{\cal I}_{ij} x^i x^j}\over
r^5} - {\cal E}_{ij} x^i x^j ,  \label{eq:h00} \\
h_{0j} &\equiv A_j = - 2 {{\dot {\cal I}}_{ja} x^a \over r^3}
                  - {10 \over 21} {\dot{\cal E}}_{ab} x^a x^b x^j
                  + {4 \over 21} {\dot{\cal E}}_{ja} x^a r^2 , \label{eq:h0j} \\
h_{ij} &= -2\Phi \delta_{ij} = \delta_{ij} (2 {M \over r} 
            + 3 {{{\cal I}_{ij} x^i x^j}\over r^5}
            - {\cal E}_{ij} x^i x^j) ,  \label{eq:hij}
\end{align}
where $\Phi$ is a scalar potential analogous to that of Newtonian gravity and $A_j$ is a vector potential that has no Newtonian analog.

This metric, by virtue of the de Donder gauge conditions and the approximations mentioned above, satisfies the following relations, which we shall use in our calculations below:
\begin{align}
A_{j,0} &= 0 , \label{deDondercondition1} \\
A_{j,j} &= -4\Phi_{,0} , \label{deDondercondition2}\\
{h_{\mu \nu ,\sigma}}^{ \sigma} &= 0 ,\label{hcommacomma1} \\
{h_{,\sigma}}^{\sigma} &= 0 .  \label{hcommacomma2}
\end{align}

To compute the quadratic part of the metric $k_{\alpha \beta}$ one would have
to solve the second order Einstein field equations. It turns out that for the
pseudotensors we are considering, a suitable choice of gauge will make the
direct calculation of $k_{\alpha \beta}$ unnecessary. This specific gauge will
be discussed later.
 
Other formulas that will be useful in the calculations that follow are given
below, accurate up to order $\varepsilon^{2}$:
\begin{equation}
g^{\alpha \beta} = \eta^{\alpha \beta} - \varepsilon h^{\alpha \beta} +
\varepsilon^{2} (h^{\alpha \gamma} h^{\beta}_{\, \gamma}) - \varepsilon^{2}
k^{\alpha \beta} ,  \label{gup}
\end{equation}
\begin{equation}
g^{\alpha \beta} g_{\gamma \beta} = \delta^{\alpha}_{\gamma} +
O(\varepsilon^{3}) , \label{gginv} 
\end{equation}
\begin{equation}
-g = {\text{det}}(g_{\alpha \beta})=1 + \varepsilon h + {1 \over{2}}
\varepsilon^{2}(h^{2} - h^{\alpha \beta} h_{\alpha \beta}) + \varepsilon^{2} k
. 
\label{detg}
\end{equation}
Taylor expanding Eq. (\ref{detg}) about $\varepsilon$, we also have, accurate to order $\varepsilon^{2}$,
\begin{equation}
\sqrt{-g}=1 + {1 \over{2}} \varepsilon h + {1 \over{2}} \varepsilon^{2}({1
\over{4}} h^{2} - {1 \over{2}} h^{\alpha \beta} h_{\alpha \beta}) + {1
\over{2}} \varepsilon^{2} k ,
\label{detsqrtg} 
\end{equation}
\begin{equation}
{1\over{\sqrt{-g}}} = 1 - {1 \over{2}} \varepsilon h + {1 \over{2}}
\varepsilon^{2}({1 \over{4}} h^{2} + {1 \over{2}} h^{\alpha \beta} h_{\alpha
\beta}) - {1 \over{2}} \varepsilon^{2} k. 
\label{detoneoversqrtg}
\end{equation} 
We are now ready to compute the tidal work using various pseudotensors.

\subsection{Calculation of Tidal Work using the Einstein Pseudotensor}
\label{subsec:tidaleinstein}

We wish to calculate the integral given by Eq. (\ref{surfaceint}) for the rate of change of mass-energy of the isolated body using the Einstein pseudotensor${\sqrt{-g}} {}_{\text{E}}{t_{0}}^{j}$: 
\begin{equation}
{{dM}\over{dt}}=- \oint {\sqrt{-g}} {}_{\text{E}}{t_{0}}^{j} n_j r^2 d\Omega , 
\label{surfaceinteinstein}
\end{equation} 
where $n_j=x_{j}/r$ is the unit normal to a surface lying in the buffer zone at
some radius $r$, and $d\Omega$ is the solid angle on that surface.
If we look at the form of $\sqrt{-g} {}_{\text{E}} {t_{\mu}}^{\nu}$ given by Eq. (\ref{tE2}), we
can see that in order to obtain an expression that is accurate up to order
$\varepsilon^{2}$ we only need to expand the metric to order $\varepsilon$.
This means that we can ignore all terms appearing in Eqs. (\ref{eq:metric}),
(\ref{gup}) -- (\ref{detoneoversqrtg}) that go as $\varepsilon^{2}$. Eq.
(\ref{tE2}) thus takes the form (in a general gauge):
\begin{multline}
\sqrt{-g} {}_{\text{E}} {t_{\mu}}^{\nu} = {\varepsilon^{2} \over{16 \pi}} \bigg\{
{1\over 2} h_{,\mu} {h^{\nu \alpha}}_{,\alpha} - {1\over 4} h_{,\mu}h^{,\nu} -
{h^{\alpha \beta}}_{,\mu}{h^{\nu}}_{\alpha ,\beta}  \\
+ {1\over 2} {h^{\alpha \beta}}_{,\mu}{h_{\alpha \beta}}^{,\nu} - {1\over 4}
h_{,\mu}h^{,\nu} + {1\over 2} h_{,\alpha} {h^{\alpha \nu}}_{,\mu} -
{\delta_{\mu}}^{\nu} \Big( {1\over 2} h_{,\gamma} {h^{\gamma \alpha}}_{,\alpha}
\\
 - {1\over 4} h_{,\gamma}h^{,\gamma} - {1\over 2} h^{\alpha \sigma
,\gamma}h_{\alpha \gamma ,\sigma} + {1 \over 4} h^{\alpha \sigma
,\gamma}h_{\alpha \sigma ,\gamma} \Big) \bigg\} .
\label{tE3}
\end{multline}
Applying the de Donder gauge constraints (\ref{dedonder}), we see that the
first and second and the seventh and eighth terms cancel each other.

From Eq. (\ref{surfaceinteinstein}) we see that we only need to evaluate the
${}_E {t_{0}}^{j}$ terms.  These
terms evaluate\footnote{\label{GRTensornote}Part of this calculation was performed using the tensor algebra package GRTensorII \cite{grtensor}.} to
\begin{equation}
\sqrt{-g} {}_{\text{E}} {t_{0}}^{j}={1\over{4 \pi}} \Phi_{,0} \Phi_{,j}.
\label{t0j}
\end{equation}
Using Eq. (\ref{eq:h00}) we see that
\begin{multline}
\Phi_{,0} \Phi_{,j} = {15\over{4}} {\cal{I}}_{ab} \dot{{\cal{E}}}_{cd} {{x_a
x_b x_c x_d x_j}\over{r^7}} \\ - {3\over{2}} {\cal{I}}_{ja}
\dot{{\cal{E}}}_{bc} {{x_a x_b x_c}\over{r^5}} - {3\over{2}}
\dot{{\cal{I}}}_{ab} {\cal{E}}_{jc} {{x_a x_b x_c}\over{r^5}}  .
\label{phi0phij}
\end{multline}
Note that we have ignored terms that go like $M \dot{M}$, $\dot{M} \cal{I}$,
$\dot{M} \cal{E}$, $\dot{\cal{I}} M$, $\dot{\cal{I}} \cal{I}$, $\dot{\cal{E}}
M$, and $\dot{\cal{E}} \cal{E}$ as they do not contribute to the tidal work.
That these terms do not contribute is apparent if one considers that the tidal
work must arise due to a coupling between the mass multipole moments of the
isolated body and the tidal field of the external universe. This, combined with
dimensional considerations, implies that only terms of the form $\dot{\cal{I}}
\cal{E}$ and $\cal{I} \dot{\cal{E}}$ can contribute to $dW/dt$.   Also we do
not worry about spatial indices being up or down since we are using Cartesian
coordinates.

We must now evaluate the surface integral given in Eq.
(\ref{surfaceinteinstein}). Note that keeping our calculations accurate to
order $\varepsilon^{2}$ justifies our setting the factor of $\sqrt{-g}$ on the
left-hand-side of Eq. (\ref{t0j}) to unity. To perform the surface integrals
over the terms in Eq. (\ref{phi0phij}), we first note that since the multipole
moments do not vary spatially in the buffer region, they can be pulled out of
the integrals. The surface integrals that remain are all of the form
\begin{equation}
\oint n_a n_b n_c \ldots n_p \, d\Omega \, ,
\label{generalsurfaceint}
\end{equation}
where $n_a=x_a/r$ is a component of a unit radial vector. Evaluating such
integrals (see Sec. IIB of Thorne \cite{thornemultipole}) we finally arrive at
\begin{equation}
{dM\over{dt}}=- {1\over{2}} {\cal{E}}_{jk} {d {\cal{I}}_{jk} \over{dt}} + {d
\over{dt}} \left( {3\over{10}} {\cal{I}}_{jk} {\cal{E}}_{jk} \right).
\label{tidalheatingeinstein}
\end{equation}

We can identify the second term in the equation above as the analog of the
Newtonian interaction-energy given by Eq. (\ref{eq:eint}) where the Einstein
pseudotensor localization corresponds to the choice $\alpha=1$. That this term
is in fact the derivative of an interaction-energy term is apparent if once
considers that the interaction energy $E_{\text{int}}$ between an isolated body
and the external universe must depend only on the instantaneous fields and can
only be given by a product $\sim {\cal{I}}_{jk} {\cal{E}}_{jk}$. The rate of
change of this interaction energy must then be a perfect differential. Also, a
term that goes like $\sim {\cal{I}}_{jk} \dot{{\cal{E}}}_{jk}$ could not
contribute to the tidal heating since no work is done if the isolated body does
not change (just as no work is done if a force is exerted but no displacement
results); see Sec. I for further discussion. These facts indicate that the first term in Eq.
(\ref{tidalheatingeinstein}) is the tidal work  while the second term is the rate of
change of the interaction energy between the external universe and the body; cf. Eq. (\ref{Decompose}).  

\subsection{Calculation of Tidal Work using the Landau-Lifshitz Pseudotensor}
\label{subsec:tidalLL}

The calculation of the tidal work using the Landau-Lifshitz pseudotensor is
very similar to that for the Einstein pseudotensor shown above. This calculation
was performed by Purdue \cite{purdue} and we will only summarize her results here.

Since the Landau-Lifshitz pseudotensor, like the Einstein pseudotensor, is
quadratic in the first derivatives of the metric, one only needs to expand
$g_{\alpha \beta}$ to first order in $\varepsilon$. We thus only need to
consider the linear part of the metric $h_{\alpha \beta}$ when evaluating the
integral in Eq. (\ref{surfaceinteinstein}) (where $ {\sqrt{-g}} {}_{\text{E}}
{t_{\mu}}^{\nu}$ is replaced by $(-g) t_{\text{LL}}^{\mu \nu}$).

Evaluating this integral and keeping only terms that contribute to the tidal
work (in the same manner as the previous section), Purdue arrives at 
\begin{equation}
{dM\over{dt}}=- {1\over{2}} {\cal{E}}_{jk} {d {\cal{I}}_{jk} \over{dt}} + {d
\over{dt}} \left( -{1\over{10}} {\cal{I}}_{jk} {\cal{E}}_{jk} \right).
\label{tidalheatingLL}
\end{equation}
We notice that changing the energy localization scheme from that of Einstein to
that of Landau-Lifshitz has simply changed the coefficient of the second term,
which we again identify as the derivative of the interaction energy. Note that
the Landau-Lifshitz localization scheme is analogous to a choice of $\alpha=
-3$ in the Newtonian interaction term. We also see that the tidal work term
(the first term in Eq. (\ref{tidalheatingLL})) has remained unaffected. 

\subsection{Calculation of Tidal Work using the M\o ller Pseudotensor}
\label{subsec:tidalmoller}

We shall now perform the calculation of $dM/dt$ once again,
this time making use of the M\o ller pseudotensor. Since we are working in the
vacuum buffer zone where $T^{\mu \nu}=0$, we can use Eq.
(\ref{mollerpseudotensor}) as the expression for the energy-momentum
pseudotensor of the gravitational field. Note that Eq.
(\ref{mollerpseudotensor}) is actually the total conserved complex that we
would use if non-gravitational fields were also present.

If we examine closely the form of Eq. (\ref{mollerpseudotensor}) we will find
that unlike the two previous pseudotensors discussed, the M\o ller complex has
a term that is linear in the second derivatives of the metric perturbations.
This means that we will not only have terms like ${h^{\alpha \beta}}_{,\mu}
{h_{\alpha \beta}}^{,\nu}$ but will also have terms like
${{{h_{\mu}}^{\sigma}}_{,\sigma}}^{\nu}$ and
${{{k_{\mu}}^{\sigma}}_{,\sigma}}^{\nu}$. This means that it is important that
we expand the metric up to quadratic order in $\varepsilon$ as these terms
that are linear in the second derivatives of the metric perturbation are
actually quadratic in $\varepsilon$ and will thus contribute to the calculation
of the tidal work.    

Using the metric given by Eq. (\ref{eq:metric}) in Eq.
(\ref{mollerpseudotensor}), we arrive at the form of the M\o ller total
energy-momentum pseudotensor, correct up to order $\varepsilon^{2}$, in a general gauge:
\begin{multline}
- 8\pi {}_{\text{M}}{\mathfrak{T}_{\mu}}^{\nu} = \varepsilon \big(
{{{h_{\mu}}^{\sigma}}_{,\sigma}}^{\nu} - {{{h_{\mu}}^{\nu}}_{,\sigma}}^{\sigma}
\big) + \varepsilon^{2} \big( {{{k_{\mu}}^{\sigma}}_{,\sigma}}^{\nu} -
{{{k_{\mu}}^{\nu}}_{,\sigma}}^{\sigma} \big) \\
 + {1\over{2}} \varepsilon^{2} \big( h_{\sigma} {h_{\mu}}^{\sigma ,\nu} + h
{{{h_{\mu}}^{\sigma}}_{,\sigma}}^{\nu}  - h_{,\sigma} {h_{\mu}}^{\nu ,\sigma} -
h {{{h_{\mu}}^{\nu}}_{,\sigma}}^{\sigma}  \big) \\
 + \varepsilon^{2} \big( {h^{\sigma \alpha}}_{,\sigma}
{{h_{\mu}}^{\nu}}_{,\alpha} + h^{\sigma \alpha} {{h_{\mu}}^{\nu}}_{,\alpha
\sigma}  - {h^{\sigma \alpha}}_{, \sigma} {h_{\mu \alpha}}^{,\nu} - h^{\sigma
\alpha} {{h_{\mu \alpha}}_{,\sigma}}^{\nu} \\
+  {h^{\nu \beta}}_{,\sigma} {h_{\mu \beta}}^{,\sigma} +  h^{\nu \beta}
{{h_{\mu \beta}}_{, \sigma}}^{\sigma} -{h^{\nu \beta}}_{,\sigma}
{{h_{\mu}}^{\sigma}}_{,\beta}  - h^{\nu \beta} {{h_{\mu}}^{\sigma}}_{,\beta
\sigma}  \big). \\ 
\label{moller2ndorder}
\end{multline}
If we work in de Donder gauge to linear order we can use Eqs. (\ref{eq:h00}) -- (\ref{eq:hij}) for $h_{\alpha \beta}$ . However, we still do
not know the form of $k_{\alpha \beta}$. 

Fortunately, we can make use of the vacuum
Einstein field equations 
\begin{equation}
R_{\mu \nu}={\Gamma^\alpha}_{\mu \nu ,\alpha} - {\Gamma^\alpha}_{\mu \alpha ,
\nu} + {\Gamma^\alpha}_{\beta \alpha} {\Gamma^\beta}_{\mu \nu} -
{\Gamma^\alpha}_{\beta \nu} {\Gamma^\beta}_{\mu \alpha} = 0
\label{ricci}
\end{equation} 
to solve for the derivatives of $k_{\alpha \beta}$ as they appear in Eq.
(\ref{moller2ndorder}).  Specifically: If we substitute the metric (\ref{eq:metric}) in Eq.
(\ref{ricci}) we can expand the Ricci tensor in powers of $\varepsilon$:
\begin{equation}
\varepsilon R^{(1)}_{\mu \nu}[h] + \varepsilon^{2} R^{(2)}_{\mu \nu}[hh] +
\varepsilon^{2} R^{(2)}_{\mu \nu}[k] = 0 ,
\label{riccipower}
\end{equation}
where the superscript on the $R_{\mu \nu}^{(n)}$ means that the indicated piece of the Ricci tensor contains only terms of
order $\varepsilon^{n}$. The terms in the brackets indicate that the part of
the Ricci tensor in question contains terms that go like the indicated multiple
of the metric piece ($h_{\alpha \beta}$ or $k_{\alpha \beta}$) and its
derivatives. Eqs. (\ref{1storderefe}) and (\ref{2ndorderefe}) make this clear.
 We now require that the vacuum field equations vanish in each order of
$\varepsilon$:
\begin{equation}
R^{(1)}_{\mu \nu}[h]=0\,,
\label{R1st}
\end{equation}
\begin{equation}
R^{(2)}_{\mu \nu}[hh] +  R^{(2)}_{\mu \nu}[k] = 0 .
\label{R2nd}
\end{equation}
The first of these equations yields the linearized vacuum field equations:
\begin{equation}
{h^{\alpha}}_{\mu , \nu \alpha} + {h^{\alpha}}_{ \nu , \mu \alpha} - {{h_{\mu
\nu}}_{, \alpha}}^{\alpha} - h_{, \mu \nu} = 0 \,,
\label{1storderefe}
\end{equation}
while in the second (\ref{R2nd}) one can solve for $R^{(2)}_{\mu
\nu}[k_{\alpha \beta}]$ to give:
\begin{multline}
{k^{\alpha}}_{\mu , \nu \alpha} + {k^{\alpha}}_{ \nu , \mu \alpha} - {{k_{\mu
\nu}}_{, \alpha}}^{\alpha} - k_{, \mu \nu}  = {h^{\alpha \sigma}}_{,\alpha}
h_{\sigma \mu , \nu} \\
 +  h^{\alpha \sigma} h_{\sigma \mu ,\nu \alpha} + {h^{\alpha
\sigma}}_{,\alpha} h_{\sigma \nu ,\mu} + h^{\alpha \sigma} h_{\sigma \nu ,\mu
\alpha} -  {h^{\alpha \sigma}}_{,\alpha} h_{\mu \nu , \sigma} \\
- h^{\alpha \sigma} h_{\mu \nu ,\sigma \alpha} - {1\over{2}} {h^{\alpha
\sigma}}_{,\nu} h_{\sigma \alpha ,\mu} - h^{\alpha \sigma} h_{\sigma \alpha
,\mu \nu} -{1\over{2}} h_{,\beta} {{h_{\mu}}^{\beta}}_{,\nu} \\
-{1\over{2}} h_{,\beta}{{h_{\nu}}^{\beta}}_{,\mu}  + {1\over{2}} h_{,\beta}
{h_{\mu \nu}}^{,\beta}  + {h_{\beta \nu}}^{,\alpha} {h_{\mu \alpha}}^{,\beta}
-{h_{\beta \nu}}^{,\alpha} {{h_{\mu}}^{\beta}}_{,\alpha} . \\
\label{2ndorderefe}
\end{multline}
We can now use Eq. (\ref{2ndorderefe}) to substitute for the $k_{\alpha \beta}$
terms that appear in Eq. (\ref{moller2ndorder}), if we pick a gauge for the
$k_{\alpha \beta}$ such that
\begin{equation}
{k^{\alpha}}_{\mu , \nu \alpha} + {k^{\alpha}}_{ \nu , \mu \alpha} - {{k_{\mu
\nu}}_{, \alpha}}^{\alpha} - k_{, \mu \nu} = {{k_{\mu}}^{\sigma}}_{,\nu \sigma}
- {{k_{\mu \nu}}_{,\sigma}}^{\sigma}. 
\label{relatek}
\end{equation}
This is easily done if we choose the gauge
\begin{equation}
{{k_{\nu}}^{\alpha}}_{,\alpha} = k_{,\nu}.
\label{kgauge}
\end{equation}   

We can now use Eqs. (\ref{2ndorderefe}) and (\ref{kgauge}) along with the de
Donder gauge conditions and Eqs. (\ref{hcommacomma1}) and (\ref{hcommacomma2})
to simplify the M\o ller pseudotensor (\ref{moller2ndorder}) into the form:
\begin{multline}
- 8\pi {}_{\text{M}}{\mathfrak{T}_{\mu}}^{\nu} = {1\over{2}}{h_{,\mu}}^{\nu} +
{1\over{4}} h {h_{,\mu}}^{\nu} - {1\over{2}}h_{,\mu \beta} h^{\nu \beta}  \\
+ h^{\alpha \sigma}{{h_{\sigma}}^{\nu}}_{,\mu \alpha}  -{1\over{2}} h^{\alpha
\sigma , \nu} h_{\alpha \sigma ,\mu} - h^{\alpha \sigma} {h_{\alpha \sigma
,\mu}}^{\nu} .
\label{mollersimplified}
\end{multline}
Inserting Eqs. (\ref{eq:h00}) -- (\ref{eq:hij}) for $h_{\alpha \beta}$, we obtain$^{\ref{GRTensornote}}$:
\begin{equation}
- 8\pi {}_{\text{M}}{{\mathfrak{T}}_{0}}^{j} = -2 \Phi_{,j 0} -8 \Phi_{,j}\Phi_{,0}
-12\Phi\Phi_{,j 0}.
\label{mollersimplified2}
\end{equation}
Expanding this expression and ignoring terms in the same manner as Eq.
(\ref{phi0phij}), we plug into the integral
\begin{equation}
{{dM}\over{dt}}=- \oint {}_{\text{M}}{{\mathfrak{T}}_{0}}^{j} n_j r^2 d\Omega , 
\label{surfaceintmoller}
\end{equation}
and arrive at
\begin{equation}
{{dM}\over{dt}}=-\dot{M} - {\dot{{\cal{I}}}}_{jk} {\cal{E}}_{jk},
\label{tidalmoller1}
\end{equation}
where the $\dot{M}$ term comes from the $-2 \Phi_{,j 0}$ term in
(\ref{mollersimplified2}), which in turn arises from the fact that the M\o ller pseudotensor contains a piece linear in the second derivatives of the metric. Eq. (\ref{tidalmoller1}) can then be written as
\begin{equation}
{{dM}\over{dt}}=- {1\over{2}} {\cal{E}}_{jk} {d {\cal{I}}_{jk} \over{dt}} .
\label{tidalheatingmoller}
\end{equation}
Note that in the case of the M\o ller pseudotensor, the perfect differential
term that represents the interaction energy vanishes. This serves to support
our intuition that the interaction term is simply a mathematical artifact of
our choice of energy localization and that the tidal work is in fact
uniquely given by Eq. (\ref{tidalwork}). Comparison with Eq. (\ref{eq:eint})
implies that the M\o ller pseudotensor corresponds to a Newtonian
energy localization given by $\alpha= -2$.

\subsection{Calculation of Tidal Work using the Bergmann Conserved
Quantities}
\label{subsec:tidalbergmann}

To further confirm our intuition about the tidal work, we can calculate
$dM/dt$ using the Bergmann conserved quantities given by Eq.
(\ref{strongconserve}). However, before we do this, we must determine the form of the arbitrary vector field $\xi^{\sigma}$, on which the momentum density $D^{\mu}$ depends. This form must be such that the integral (\ref{bergsurfint1}) gives
\begin{equation}
P^{0}=\varepsilon M + O(\varepsilon^{2}),
\label{constrainform}
\end{equation}
the mass-energy $M$ plus terms of order $\varepsilon^{2}$ which arise from the fact that the surface of integration $\partial {\cal{V}}$ lies in the buffer zone, where spacetime is only \emph{locally} asymptotically flat.

Computing $P^0$ via Eq. (\ref{bergsurfint1}) first involves calculating the von Freud
superpotential (\ref{vonfreudsuper}) in terms of the metric expansion given
by Eq. (\ref{eq:metric}). After a rather laborious
computation, we find the von Freud superpotential in general gauge and to
quadratic order in the metric perturbation:
\begin{multline}
- 16 \pi {}_{\text{F}} U_{\alpha}^{[\beta \gamma]} = \varepsilon \bigg\{ \big(
{\delta_{\alpha}}^{\beta} h^{,\gamma} - {\delta_{\alpha}}^{\gamma} h^{,\beta}
\big) \\
+ \big( {\delta_{\alpha}}^{\gamma} {h^{\beta \lambda}}_{,\lambda} -
{\delta_{\alpha}}^{\beta} {h^{\gamma \lambda}}_{,\lambda} \big) + \big(
{h_{\alpha}}^{\gamma ,\beta} - {h_{\alpha}}^{\beta ,\gamma} \big) \bigg\} \\
 + \varepsilon^2 \bigg\{ {1\over 2} h \big( {\delta_{\alpha}}^{\beta}
h^{,\gamma} - {\delta_{\alpha}}^{\gamma} h^{,\beta} \big) + {1\over 2} h \big(
{\delta_{\alpha}}^{\gamma} {h^{\beta \lambda}}_{,\lambda} -
{\delta_{\alpha}}^{\beta} {h^{\gamma \lambda}}_{,\lambda} \big) \\
+ {1\over 2} h \big( {h_{\alpha}}^{\gamma ,\beta} - {h_{\alpha}}^{\beta
,\gamma} \big) + \big( h^{\gamma \lambda} {h^{\beta}}_{\alpha ,\lambda} -
h^{\beta \lambda} {h^{\gamma}}_{\alpha ,\lambda} \big) \\
 + h_{,\lambda} \big( {\delta_{\alpha}}^{\gamma} h^{\beta \lambda} -
{\delta_{\alpha}}^{\beta} h^{\gamma \lambda} \big) + {h^{\lambda}}_{\rho} \big(
{\delta_{\alpha}}^{\beta}  {h^{\gamma \rho}}_{,\lambda} -
{\delta_{\alpha}}^{\gamma} {h^{\beta \rho}}_{,\lambda} \big) \\ 
+  {h^{\lambda}}_{\rho ,\lambda} \big( {\delta_{\alpha}}^{\beta} h^{\gamma
\rho} - {\delta_{\alpha}}^{\gamma} h^{\beta \rho} \big) + \big( h^{\beta \rho}
{h_{\alpha \rho}}^{,\gamma} - h^{\gamma \rho} {h_{\alpha \rho}}^{,\beta} \big)
\\
+ h_{\kappa \rho} \big( {\delta_{\alpha}}^{\gamma} h^{\kappa \rho ,\beta} -
{\delta_{\alpha}}^{\beta} h^{\kappa \rho ,\gamma} \big) + \big(
{\delta_{\alpha}}^{\beta} k^{,\gamma} - {\delta_{\alpha}}^{\gamma} k^{,\beta}
\big) \\
+ \big( {\delta_{\alpha}}^{\gamma} {k^{\beta \lambda}}_{,\lambda} -
{\delta_{\alpha}}^{\beta} {k^{\gamma \lambda}}_{,\lambda} \big) +  \big(
{k_{\alpha}}^{\gamma ,\beta} - {k_{\alpha}}^{\beta ,\gamma} \big) \bigg\} .
\label{fullvonfreud}
\end{multline}  

As in the case of the M\o ller pseudotensor, we have terms that are linear
in the derivatives of the $k_{\alpha \beta}$. For our calculation of $dM/dt$ below, we will find that these terms may again be expressed in terms of the $h_{\alpha \beta}$ by the same choice of gauge as we used in the previous section, Eq. (\ref{kgauge}).
However, we cannot do this for our present calculation of $P^0$. Fortunately, this will not be a problem since we only wish to show that $P^0$ reduces to the mass-energy $M$ plus terms of order $\varepsilon^{2}$. Since $k_{\alpha \beta}$ is of order $\varepsilon^2$ it is not necessary to include the terms containing $k_{\alpha \beta}$ in our calculation of the surface integral in Eq. (\ref{bergsurfint1}). This is also true of the terms in (\ref{fullvonfreud}) that are products of $h_{\alpha \beta}$ and its derivatives. We therefore only need concern ourselves with the first six terms in Eq. (\ref{fullvonfreud}) (those linear in $h_{\alpha \beta}$ and its derivatives) as only these terms could conceivably affect $P^0$ at order $\varepsilon$.

Now, let us make a guess as to the form of the vector field $\xi^{\sigma}$, on which $D^{\mu}$ depends. A volume integral of Bergmann's $D^{0}$ (or the equivalent surface integral of (\ref{bergsurfint1})) will reduce to Eq. (\ref{constrainform}) only if the form of $\xi^{\sigma}$ is properly constrained. If our spacetime were precisely asymptotically flat, we would expect $\xi^{\sigma}$ to be \emph{asymptotically} the timelike Killing vector field, $\xi^{\sigma} \to {{\partial}/{\partial t}} + O(1/r)$. Since our spacetime is not
asymptotically flat but only \emph{locally} asymptotically flat, it is reasonable to expect that
$\xi^{\sigma}$ is locally asymptotically Killing, by which we mean that it can
be written as the sum of a timelike vector, $\delta^{\sigma}_{0}$, plus
deviations $\zeta^{\sigma}$, from this timelike vector that are due to the
fact that the spacetime is not flat:
\begin{equation}
 \xi^{\sigma}=\delta^{\sigma}_{0} + \zeta^{\sigma}.
\label{xi1}
\end{equation}

As for the form of $\zeta^{\sigma}$, we can construct a quantity which is the
most general vector field that is a) dimensionally correct and b) constructed
only out of the parameters that characterize our spacetime in the slow-motion
approximation: $M$, ${\cal{I}}_{jk}$, ${\cal{E}}_{jk}$ and their first time
derivatives. Such a vector field has the following form (to order $\varepsilon$):
\begin{multline}
\zeta^{0} = a {\dot{M}} + b {M \over r}+ c {{\cal{I}}_{ab} x_a x_b
\over{r^{5}}} \\
+ d {{\dot{\cal{I}}}_{ab} x_a x_b \over{r^{4}}} + e {\cal{E}}_{ab} x_a x_b + f
{\dot{\cal{E}}}_{ab} x_a x_b r , \label{zeta0}
\end{multline}
\begin{multline}
\zeta^{k} = g {M x_k \over{r^2}} + h {{\dot{M}} x_k \over{r^2}} + i
{{\cal{I}}_{ab} x_a x_b x_k \over{r^6}} 
+ j {{\dot{\cal{I}}}_{ab} x_a x_b x_k \over{r^5}} \\
+ k {{\cal{I}}_{ak} x_a \over{r^4}} + l {{\dot{\cal{I}}}_{ak} x_a \over{r^3}} +
m {{\cal{E}}_{ab} x_a x_b x_k \over{r}} \\
+ n {\dot{\cal{E}}}_{ab} x_a x_b x_k + o {\cal{E}}_{ak} x_a r + p
{\dot{\cal{E}}}_{ak} x_a r^2 , \label{zetaj}
\end{multline}
where the coefficients $a$ through $p$ are real-valued constants.

Using this prescription, we are now able to evaluate $P^0$. Our result reduces to Eq. (\ref{constrainform}) as required. We thus see that the surface integral (\ref{bergsurfint1}) does indeed give the mass as the only $O(\varepsilon)$ term,  plus terms of $O(\varepsilon^2)$ that arise from the facts that our spacetime is \emph{locally} asymptotically flat and that we evaluate $P^0$ on the 2-surface $\partial {\cal{V}}$ that lies at some finite $r$ in the buffer zone.

Now we wish to compute the tidal work, $dP^{0}/dt$ or $dM/dt$. Taking a time derivative of Eq. (\ref{bergint}) and applying Gauss' law, we easily arrive at the expression
\begin{equation}
{dM \over{dt}} = - \oint \left( \xi^{\sigma} {}_{\text{F}} U_{\sigma}^{[j \nu]}
\right) _{,\nu} \, d^2S_j.
\label{dpdt}
\end{equation}
 Plugging Eq. (\ref{xi1}) into Eq. (\ref{dpdt}) allows us to write
\begin{equation}
{dM\over{dt}}=-\oint {}_{\text{F}} {U_0 ^{[j \nu]}}_{,\nu} \, d^2S_j - \oint \left(
\zeta^{\sigma} {}_{\text{F}} U_{\sigma} ^{[j \nu]} \right) _{,\nu} d^2S_j .
\label{dpdt2}
\end{equation}

Let us now think carefully about what terms we actually need to calculate in
these integrals. If we examine the first integral in Eq. (\ref{dpdt2}) we
realize (comparing with Eq. (\ref{fullvonfreud})) that all of the terms in the integrand are of order $\varepsilon^{2}$ and can therefore contribute to the tidal
work. However, we again have the problem that we do not know the explicit form of $k_{\alpha \beta}$. Fortunately, if we
expand the terms in $ \bigl({}_{\text{F}} {U_{\sigma} ^{[\mu \nu]}} \bigr)_{,\nu}$ that
depend on $k_{\alpha \beta}$ and apply the gauge condition of Eq. (\ref{kgauge}), we find that these terms reduce to 
\begin{equation}
  \left( {{k^{\nu}}_{\sigma , \nu}}^{\mu} - {{k^{\mu}}_{\sigma ,\nu}}^{\nu}
\right).
\label{bergks}
\end{equation}   
Comparison with Eq. (\ref{relatek}) shows that this is the same situation
encountered with the M\o ller pseudotensor. We can thus use Eq. (\ref{2ndorderefe}) to
express the first integrand in Eq. (\ref{dpdt2}) entirely in terms of
$h_{\alpha \beta}$ and its derivatives, just as we did in the previous section. Applying de Donder gauge and a fair
amount of algebraic manipulation we arrive at 
\begin{multline}
-16 \pi {}_{\text{F}} {U_{\sigma}^{[\mu \nu]}}_{,\nu} = {1\over 2} h_{,\lambda} {h^{\mu
\lambda}}_{,\sigma} - {1\over 4} h_{,\sigma} h^{,\mu} + {\delta_{\sigma}}^{\mu}
h^{\lambda \rho ,\nu} h_{\rho \nu ,\lambda} \\
- h_{\lambda \rho ,\sigma} h^{\mu \rho ,\lambda} + {1\over 2} h^{\lambda \rho
,\mu} h_{\lambda \rho ,\sigma} - {\delta_{\sigma}}^{\mu} h^{\kappa \rho ,\nu}
h_{\kappa \rho ,\nu}.
\label{bergsimp}
\end{multline}
Evaluating the components we need for the first integral in Eq. (\ref{dpdt2}), we find$^{\ref{GRTensornote}}$
\begin{equation}
{}_{\text{F}} {U_0 ^{[j \nu]}}_{,\nu} = - {1\over{4\pi}} \Phi_{,0} \Phi_{,j}.
\label{bergsimp2}
\end{equation}
If we expand this expression as we did in Eq. (\ref{phi0phij}), keeping only
terms that can possibly contribute to the tidal work (the products of
$\dot{\cal{I}} {\cal{E}}$ and ${\cal{I}} \dot{\cal{E}}$),  and evaluate the first integral in Eq. (\ref{dpdt2}), we
have
\begin{equation}
-\oint {}_{\text{F}} {U_0 ^{[j \nu]}}_{,\nu} \, d^2S_j = - {1\over{2}} {\cal{E}}_{jk} {d
{\cal{I}}_{jk} \over{dt}} + {3\over{10}} {d \over{dt}} \left( {\cal{I}}_{jk}
{\cal{E}}_{jk} \right). 
 \label{dpdt3}
\end{equation}
One should note that this is precisely what we obtained using the Einstein pseudotensor (Eq. (\ref{tidalheatingeinstein})), and indeed, the left-hand side of (\ref{dpdt3}) is the surface integral that one would obtain by substituting Eq. (\ref{divvonfreudsuper}) into Eq. (\ref{surfaceinteinstein}).

We must now evaluate the second integral in Eq. (\ref{dpdt2}) to see if it will
contribute to the tidal work. The computation of this integral is made much
simpler if we realize that $\zeta^{\sigma}$ is linear in $\varepsilon$. Since
the tidal work is quadratic in $\varepsilon$ we only have to concern
ourselves with the piece of ${}_{\text{F}} U_{\alpha}^{[\beta \gamma]}$ that is linear
in $\varepsilon$. Applying de Donder gauge to the linear piece of Eq.
(\ref{fullvonfreud}) we can write the second integral in Eq. (\ref{dpdt2}) as
\begin{multline}
- \oint \left( \zeta^{\sigma} {}_{\text{F}} U_{\sigma} ^{[j \nu]} \right) _{,\nu} d^2S_j
= \\
{1\over{16\pi}} \oint {1\over 2} {\zeta^j}_{,\nu} h^{,\nu} - {1\over 2}
{\zeta^{\nu}}_{,\nu} h^{,j} + {\zeta^{\sigma}}_{,\nu} ({h_{\sigma}}^{\nu ,j}
-{h_{\sigma}}^{j ,\nu} ) \, d^2S_j \\
+ O(\varepsilon^3),
\label{dpdt4}
\end{multline}
where we have applied Eqs. (\ref{hcommacomma1}) and (\ref{hcommacomma2}). A few
pages of algebra, and again ignoring terms that cannot contribute to the tidal
work, reduces this to
\begin{equation}
- \oint \left( \zeta^{\sigma} {}_{\text{F}} U_{\sigma} ^{[j \nu]} \right) _{,\nu} d^2S_j
= \left( {2\over{15}} c + {3\over{5}} e \right) {d\over{dt}} ({\cal{I}}_{jk}
{\cal{E}}_{jk} ). 
\label{dpdt5}
\end{equation}

Combining Eqs. (\ref{dpdt3}) and (\ref{dpdt5}) we finally arrive at an expression for $dM/dt$ in terms of the products that can contribute
to the tidal work ($\dot{\cal{I}} \cal{E}$ and $\cal{I} \dot{\cal{E}}$):
\begin{equation}
{{dM}\over{dt}}=- {1\over{2}} {\cal{E}}_{jk} {d {\cal{I}}_{jk} \over{dt}} +
\left( {3\over{10}} + {2\over{15}} c + {3\over{5}} e \right) {d\over{dt}}
\left({\cal{I}}_{jk} {\cal{E}}_{jk} \right). 
\label{tidalberg}
\end{equation}

We can now see conclusively that changing the energy localization has no effect
on the tidal work term and merely changes the coefficient in front of the
arbitrary interaction-energy term.

\section{Conclusions}
This paper completes a demonstration that the tidal work caused by the interaction of an isolated body's quadrupole moment ${\cal{I}}_{jk}$ with the electric-type tidal field ${\cal{E}}_{jk}$ of an external universe is unambiguous, despite the ambiguity in the definition of the mass $M$ of such a system. Purdue \cite{purdue} demonstrated that a gauge change does not lead to any ambiguity in the tidal work in general relativity, nor does one's choice of energy localization in Newtonian theory. We have shown that this energy localization invariance carries over into the general relativistic description. In addition, the work of Booth \& Creighton \cite{jolien}, carried out simultaneous with our own, supports the conclusions of Purdue and ourselves through an independent, though equivalent, approach using the techniques of quasilocal energy.

One of the main motivations for this paper was to strengthen the arguments used by Thorne \cite{thornetidal} in his analysis of the stability of neutron stars against radial collapse induced by an external tidal field. The motivation for Thorne's paper was to refute the claims of Wilson, Mathews, and Marronetti \cite{wmm} regarding the ``star-crushing'' effect seen in their numerical simulations of binary neutron stars. We believe that our analysis has strengthened Thorne's arguments. Ironically, Thorne and I \cite{favatathorne} now have reason, based on \emph{current} quadrupole tidal coupling, to support the ``star-crushing'' effects observed by Wilson and Mathews \cite{revisedwmm} in their revised simulations. 

This analysis has only been concerned with the tidal work involving a mass quadrupole moment interacting with an electric-type, quadrupolar tidal field in the slow motion approximation. It seems quite likely that this energy localization and gauge invariance of tidal work can be extended to higher order multipolar couplings. We will leave the demonstration of this for the future.

\section{Acknowledgments}
I thank Kip Thorne for suggesting this research project and for many insightful
discussions as to its solution and the prose of this paper. I also acknowledge
Patricia Purdue for useful discussions related to her paper.  This research was supported by
the Caltech Summer Undergraduate Research Fellowship (SURF) office, by NSF
grant AST-9731698, NASA grant NAG5 6840, and the generous tuition payments of my parents, Vincent and Diana Favata.

\end{document}